\documentclass[aps,amssymb,prb,showpacs,twocolumn,groupedaddress]{revtex4}
\usepackage{graphicx}
\usepackage{dcolumn}
\usepackage{bm}
\usepackage{color}

\begin{document}

\preprint{APS,feyn54/123-QED}

\title{Enhanced angular overlap model for f-electron non-metallic systems}
\author{Z. Gajek}
\email[e-mail: ]{gajek@int.pan.wroc.pl}
 \affiliation{ Institute of Low Temperature and Structure
 Research,Polish Academy of Sciences, P Nr 1410, 50-950 Wroc{\l}aw 2, Poland}
\date{\today}
\begin{abstract}
An efficient method of interpretation of the crystal field effect
in non-metallic f-electron systems, the {\it enhanced angular
overlap model} (EAOM), is presented. The method is established on
the ground of perturbation expansion of the effective Hamiltonian
for localized electrons and first principles calculations related
to available experimental data. Series of actinide compounds,
$An$O$_2$, oxychalcogenides, $An$O$X$ and dichalcogenides U$X_2$
where $X$ = S, Se, Te and $An$ = U, Np serve as a probe of
effectiveness of the proposed method. An idea is to enhance the
usual angular overlap model with {\it ab initio} calculations of
those contributions to the crystal field potential, which cannot
be represented by the usual angular overlap model (AOM). The
enhancement leads to an improved fitting and makes the approach
intrinsically coherent. In addition, the {\it ab initio}
calculations of the main, AOM-consistent part of the crystal field
potential allows one to fix the material-specific relations for
the EAOM parameters in the effective Hamiltonian. In consequence,
the electronic structure interpretation based on EAOM can be
extended to the systems of the lowest point symmetries or/and
deficient experimental data. Several examples illustrating the
promising capabilities of EAOM are given.
\end{abstract}



\maketitle

\section{\label{intr}Introduction}

A list of phenomena traditionally discussed in the context of the
crystal field (CF) like optical excitations, Schottky effect, Van
Vleck susceptibility etc has been extended in last decades to
include so intriguing and different manifestations of many-body
effects as unconventional superconductivity, Kondo-like behavior,
magnetic rearrangements or non-Fermi liquid. Complexity of these
phenomena contrasts with apparent simplicity the one-electron,
local CF potential in the effective Hamiltonian. Nevertheless, its
reliable determination meets serious difficulties even today, 75
years after Bethe considered the CF effect for the first
time.\cite{b29} These concern not only calculations from first
principles but also common phenomenological schemes based
exclusively on the symmetry arguments. In practice, only compounds
of the highest symmetries can be handled satisfactorily. In most
cases, the usual least squares fitting of all the parameters in
the effective Hamiltonian requires additional, more or less
heuristic reasoning referring to physical or chemical foundations.
Alternatively, simplified phenomenological models are employed to
circumvent the problem of overnumerous parameters. Accuracy of
these models is one of the issues the present discussion
addresses. We focus on the angular overlap model (AOM) inspired by
the molecular orbital theory\cite{jps63,sj65} and confirmed by
further theoretical calculations.\cite{ghw81,gm92} AOM assumes,
similarly as the superposition model (SPM)\cite{n71}, the total CF
potential in the form of superposition of axial potentials due to
ligands represented by a set of intrinsic parameters. The number
of free parameters is remarkably reduced in comparison with the
basal parametrization, especially for low-symmetry systems but the
approximations are rather crude. A more refined analysis of the
experimental data, having fundamental meaning for subtle magnetic
properties or cooperative phenomena at low temperatures, may
require more rigorous methods. Can these simplified CF models be
improved without increasing the number of free parameters? This
question is discussed here on the grounds of {\it ab initio}
calculations for a series of actinide compounds.

There are two general ways to determine the CF potential from
first principles: one based on the perturbation theory and the
idea of the effective Hamiltonian for localized, open shell
electrons\cite{ghw81} and second developed on the grounds of the
density functional theory (DFT).\cite{ks65,f95} Taking into
account proximity of the band states or/and widening of the
localized {\it nf} states, DFT seems to be more suitable method
for metallic systems. However, the early implementations of DFT
based on the local spin density approximation (LSDA) had failed to
predict not only the subtle magnetic, low temperature
thermodynamic properties and the low energy spectra but even some
of the main characteristics of the crystals. For instance,
antiferromagnetic semiconductor UO$_2$ becomes a ferromagnetic
metal.\cite{k02} Over the years some inherent shortcomings of
LSDA, like double counting of states or requirement for the
electron density to be a slowly varying function, were lessened or
removed by various corrections: generalized gradient
approximation, U approximation, self-interaction correction or,
more recently, the hybrid density-functional theory.\cite{k02}
Nevertheless, these improvements, to our best knowledge, have not
eliminated completely the problems related to the CF effect (see
discussion in Ref. \onlinecite{mg00}, p. 206).

An alternative approach,\cite{gmf87,mg00} based on the
perturbation expansion of the effective Hamiltonian seems to be
more efficient and reliable in providing material-specific details
of the electronic structure. The model was developed for
non-metallic systems successively by Sugano and
Schulman,\cite{ss67} Newman and coworkers,\cite{n71,nn00} Faucher
and Garcia,\cite{fg82} and others.(see Refs. \onlinecite{nn00,
mg00} and references therein) The effective CF potential appears
in this approach as a sum of several contributions, most of which
obey the assumptions of the mentioned simplified phenomenological
models.

In metallic systems, additional mechanisms, apart from those
characteristic for insulating crystals, have to be
considered:\cite{mg00} mixing of the localized and band states
(hybridization), static and dynamic screening of the conduction
electrons including an offset screening of the carriers occupying
a virtual bond state. None of them can be regarded as a pure
superposition of the nearest neighbor contributions. Moreover, the
role of each mechanism remains unclear; for instance, the
hybridization term alone can represent almost the whole CF effect
for some compounds,\cite{lz89} but for others it is only one of
the important contributions.\cite{gfkk85} In addition, the dynamic
correlations become increasingly important as the localized state
nears the conduction band giving rise to the many-electron crystal
field effects or even leading to a breakdown of the effective
Hamiltonian theory for localized electrons. Thus, an extension of
the crystal field theory towards self-consistent models for mixed
systems of localized and itinerant electrons attracts growing
attention recently.\cite{b03} However, material-specific results
based on these efforts has not been achieved yet and we have to
confine discussion to simpler systems in which an admixture of the
band states to the localized ones or a mixing of the states
localized on adjacent atoms can be treated as an additional
perturbation term in our effective Hamiltonian. We can only notice
that there exists a subgroup of metallic systems for which the
basic CF mechanisms remain essentially the same as those for ionic
crystals.\cite{nn86}

Accuracy of the simplified models has been discussed on the
grounds of numerical simulations for actinide
compounds.\cite{gm92} It has turned out that the lattice
contributions were not always negligible and the $e_\delta$
contribution (see section \ref{f.angu}) has not behaved as a
characteristic, intrinsic parameter of the metal-ligand ($ML$)
interaction. Moreover, the present paper shows that the main AOM
energies, e$_\sigma$ and e$_\pi$, may also fail a more rigorous
test of their transferability. Thus we propose a
quasi-phenomenological approach that links the simplified
phenomenological model with partial {\it ab initio} calculations:
the {\it enhanced angular overlap model} (EAOM).

The paper is organized as follows: the perturbation expansion
leading to the effective Hamiltonian and the angular overlap model
is outlined in the section \ref{form}. Details of the {\it ab
initio} calculations, their results obtained for the first time
for U$X_2$ ($X$ = S, Se, Te)\cite{heidelberg} and those reported
previously for $An$O$_2$\cite{gfkk85} and $An$O$X$\cite{g00,g04}
are discussed in section, \ref{angu} including their reliability,
adequacy of the superposition approximation and variation of the
intrinsic parameters with distance across the series. The enhanced
angular overlap model is considered in section \ref{enha} together
with examples of its application. Concluding remarks are provided
in section \ref{conc}. To ensure self-consistency of the
presentation some known formulas used in the calculations are
included in Appendices.

\section{\label{form}Formulation}
\subsection{\label{f.effe}Effective Hamiltonian}
Since the theoretical model applied here has been presented
elsewhere (see Ref. \onlinecite{mg00} and references therein), we
only recap its main points for clarity of further discussion. The
basic assumption of the model is that all magnetic electrons in a
crystal occupy stationary orbitals, $5f$ in the case of the
discussed here actinide ions. The initial infinite many-electron
problem is reduced then to a single cluster consisting of a metal
ion and nearest neighbors - ligands. The outside of the cluster is
represented by the classic electrostatic potential. The cluster
itself is treated as a system of weakly interacting groups of
electrons localized on different ions. This allows one to apply
the group product function formalism and reduce the initial
N-electron system to several subsystems of lower
dimensions.\cite{ms69,ghw81,mg00} The zero-order group product
wave-functions are built up from the free-ion spin-orbitals
obtained with the standard self-consistent Dirac-Slater procedure
and stabilization potential wall for anions, depth of which is
determined by the Madelung energy. The function basis is
restricted to the ground and the most important inter- and
intra-ion excited electronic configurations. Projection of the
initial function space onto the ground configuration subspace,
contraction of the closed shell states, and renormalization due to
the nonorthogonality of the wave-functions centered on different
ions are the main steps in this approach. They lead to an
effective Hamiltonian defined in the restricted wave-function
space spanned by the single $5f^n$-configuration states. The
effective Hamiltonian contains several renormalization terms which
can be regarded as a perturbation to the initial Hamiltonian. The
theory may be easily verified by the experimental data because the
same function basis is employed in the conventional
phenomenological description of the electronic structure (see
Appendix A). A non-spherical part of the effective Hamiltonian
defines the effective CF potential $V({\bf r})$ "seen" by a
$5f$-electron. Since all the wave-functions in our restricted
function basis have the same radial part $\frac{1}{r}P(r)$, the
Hamiltonian and all relevant operators can be contracted to the
angular coordinates. The resulting operator of the CF potential
$\hat{V}({\bf r}/r)$ is commonly written in a form of expansion in
terms of the normalized spherical harmonics
$\hat{C}_{q}^{(k)}$,\cite{w65}
\begin{equation}\label{1}
\hat{V}({\bf r}/r)=\sum_{k,q}B_{kq}\hat{C}_{q}^{(k)}({\bf r}/r) ,
\end{equation}
where $k  = 2, 4, 6$ denotes rank of the spherical harmonic and $q
= -k, -k + 1 ,..., k$ runs over its components. $B_{kq}$ are the
integrals,
\begin{equation}
B_{kq}=\int [\frac{1}{r}P_{nf}(r)]^{2}V({\bf r})\hat{C}_{q}^{(k)\ast}%
({\bf r}/r)d{\bf r} \label{2},
\end{equation}
playing a role of adjustable CF parameters in the phenomenological
theory.
\subsection{\label{f.angu}Angular overlap model}
Angular overlap model (AOM) is a simplified phenomenological
approach based on certain restrictive assumptions, inspired by the
H\"{u}ckel molecular orbital model.\cite{jps63,sj65} From among
various formulations appearing in the literature we chose one
proposed by Sch\"{a}ffer\cite{s68} which does not refer to
molecular orbital scheme. This formulation is consistent with the
perturbation approach outlined above and the Newman superposition
model\cite{n71} discussed later. According to it, the CF potential
$V$ is a superposition of the independent contributions -
potentials $v_{t}$ generated by the nearest neighbors - ligands:
\begin{equation}
V({\bf r})\simeq\sum_{t}v_{t}(\bf{r-R}_{\it t}) \label{sup},%
\end{equation}
where $\bf R_{\it t}$ denotes the position of the $t$-th ligand.
Additionally, as an approximation, the local symmetry of the
separated metal-ligand system is assumed to be axial.

The AOM parameter $e_{\mu}^{t}$ of the given ligand $t$ is defined
as a matrix element of the ligand potential $v_{t}$ in the
coordinate system $t$ with the z axis along the metal-ligand $t$
\textit{linear ligator} in which $v_{t}$ is diagonal:
%
\begin{equation}
e_{\mu}^{t}\equiv e_{\mu}^{t}(R_{t})=\left\langle\pm\mu\left|
v_{t}\right| \pm\mu\right\rangle _{t} , \label{m2e}%
\end{equation}
where the index $\mu = 0(\sigma), 1(\pi), 2(\delta), 3(\phi)$
denotes the absolute value of the magnetic quantum number of the
$5f$-electron. It is convenient to fix the energy scale by setting
$e_{\phi}=0$. In practice we put $\widetilde{e}_{\mu}\equiv
e_{\mu}-e_{\phi}$ instead of $e_{\mu}$. Since only
$\widetilde{e}_{\mu}$'s are used hereafter, we omit the tilde for
convenience. Transformation properties under the {\it rotation
group} $R_{3}$ of the $l=3$ wave-functions allow one to express
the matrix elements of $V$ in the global coordinate system in
terms of $e_{\mu}^{t}$'s:
\begin{equation}
\left\langle m\left|  V\right| m^{\prime}\right\rangle
=\sum_{t}\sum_{\mu}D_{\mu
m}^{(3)\ast}(0,\Theta_{t},\Phi_{t})D_{\mu m^{\prime
,}}^{(3)}(0,\Theta_{t},\Phi_{t})e_{\mu}^{t} , \label{e2v}%
\end{equation}
where $D_{\mu m}^{(3)}(0,\Theta_{t},\Phi_{t})$ is the matrix
element of the irreducible representation $D^{(3)}$ of the
rotation group and $R_{t},\Theta_{t},\Phi_{t}$ are the angular
(global) coordinates of the ligand $t$. Eq. (\ref{e2v}) is the
fundamental equation of AOM. It relates the matrix elements of the
CF potential to the intrinsic parameters describing the individual
metal-ligand pairs through rotation matrices dependent on the
geometry of the coordination polyhedron.

For practical purposes it is advisable to relate the AOM
parameters to the basic CF parameters $B_{kq}$. Comparing the
right side of Eq. (\ref{e2v}) with the matrix elements of the
potential given by the expansion (\ref{1}) and using the
properties of tensor operators and the rotation
matrices\cite{e57,rbmw59} we obtain after some manipulation the
following relation:
\begin{equation}
B_{kq}=\sum_{\mu}W_{kq}^{\mu}e_{\mu}
\label{bkqem}%
\end{equation}
where
\begin{eqnarray}
W_{kq}^{\mu}=\frac{2k+1}{7}\left[  \left(
\begin{array}
[c]{ccc}%
3 \!& k\! & 3\!\\
0\! & 0\! & 0\!
\end{array}
\right)  \right]^{-1} (-1)^{\mu}(2\!-\!\delta_{\mu0})\times\nonumber\\
\times\left(
\begin{array}
[c]{ccc}%
3 \!& k\! & 3\!\\
-\mu\! & 0\! & \mu\!
\end{array}
\right)\sum_t C_{q}^{(k)\ast}(\Theta_{t},\Phi_{t}) s_{\mu}^t  \label{wkq}%
\end{eqnarray}
\begin{equation}
s_{\mu}^t=\frac{e_{\mu}^{t}}{e_{\mu}} \label{e2e}%
\end{equation}
and $(:::)$ are the $3j$ symbols. The $e_{\mu}$ parameters are the
mean values of the AOM parameters averaged over $t$. Their
introduction into Eq. (\ref{bkqem}) is one of the possible
solutions of the problem of several sets of intrinsic parameters
in the case of non-equivalent ligands. Note, that if
$W_{kq}^{\mu}$ and $s_{\mu}^t$ given by Eqs. (\ref{wkq}-\ref{e2e})
are inserted into Eq. (\ref{bkqem}), then the averaged $e_{\mu}$'s
cancel out. Moreover, the ratios $s_{\mu}^t$ can be treated as
parameters of the model instead of $e_{\mu}^t$'s. From the
algebraic point of view, this is only a scaling of the parameters
without loss of generality. In practice, it is possible to
estimate the ratios independently (this question is discussed
later) and to consider the averaged quantities $e_{\mu}$ to be
adjustable parameters. The $W_{kq}^{\mu}$ coefficients absorb all
information about the geometry of the coordination polyhedron
whereas the ratios $s_{\mu}^t$ encode differences in the AOM
parameters due to the individual $ML_{t}$ distances. The distance
dependence of $s_{\mu}^t$ has the exponential
character\cite{gm92}, yet, within a limited range of distances, it
is fairly well approximated by the simple power function:
\begin{equation}
s_{\mu}^t=\left(\frac{R}{R_t}\right)^{\alpha_{\mu}} \label{r2r}%
\end{equation}
with the power exponents $\alpha_{\mu}$ taking values in the range
4.3 to 8.9 in the case of actinide ions and simple
ligands.\cite{gm92,mg00,g00}

An extension of the above equations to arbitrary number of
different anions is straightforward. Note that the partitioning
(\ref{sup}) neglects the contribution from the outside of the
coordination cluster. This contribution and also other effects,
which cannot be represented in the form of decomposition
(\ref{sup}), are included explicitly in the enhanced model
presented below.

AOM reduces the number of adjustable parameters describing the CF
effect to three. The strength of the model stems from the fact
that the local inter-ion interaction parameters, $e_{\mu}$, may be
regarded as quantities characteristic for a given metal-ligand
($ML_{t}$) pair. They allow one to verify unphysical solutions
generated by false minima of the fitting procedures in the
standard parametric analysis on the one hand and to indicate the
ensuing approximations in the case of especially complex systems,
on the other. Due to the dominant character of the renormalization
terms (see section \ref{f.contributions} and Eq. (\ref{ren})), the
AOM parameters $e_{\sigma}$ and $e_{\pi}$ manifest several
characteristic properties:\cite{ghw81,fgj86,gm92,nn00,mg00,g00}

(i) their values reflect the spectrochemical ordering of the
anions and

(ii) decrease slightly with increasing atomic number along the
lanthanide and actinide series and, independently, with decreasing
oxidation number of the metal ion,

(iii) $e_{\sigma}>e_{\pi}>\left| e_{\delta}\right|$,

(iv) $e_{\mu}^{t}/e_{\mu}^{t'}\approx const.$ for two different
$ML_{t}$ and $ML_{t'}$ systems ($t\ne t'$),

(v) $e_{\pi}^{t}/e_{\sigma}^{t}\approx const.$ for a given
$ML_{t}$ pair,

(vi) the $e_{\delta}$ parameter, usually of minor importance, has
been shown to be 'lattice sensitive'\cite{gm92} if obtained from
the fitting of the experimental data. For this reason,
transferability of $e_{\delta}$ between various compounds seems to
be questionable in the conventional AOM.

Even though exceptions from the above rules happen (see the next
section), they may serve as an instructive test of any set of CF
parameters determined from the experimental data. This concerns
not only the fitting results obtained with other approximate
models but also the basic ${B_{kq}}$ parameters, which can be
"translated" to ${e_{\mu}}$ using Eq. (\ref{bkqem}) and a standard
least squares procedure.

As mentioned, one of the most widely applied approximate methods,
the Newman superposition model,\cite{n71,nn00} is based on the
same assumptions as AOM. However, the role of intrinsic parameters
is played by the $B_{kq}$ parameters for a separated linear
ligator in a local coordinate system. Due to the assumed
cylindrical symmetry of each $ML$ subsystem, only the parameters
with $q=0$ are effective. They are denoted hereafter as "$b_{k}$"
to distinguish them from the $B_{k0}$ parameters in the global
coordinate system. The relation between the two sets of intrinsic
parameters, AOM and SPM, can be easily obtained from Eqs.
(\ref{bkqem},\ref{wkq},\ref{e2e}) by rewriting them for the
specific case of the separated linear ligator:
\begin{eqnarray}\label{e2b}%
b_{k}  & = & \frac{2k+1}{7}\left[  \left(
\begin{array}
[c]{ccc}%
3 \!& k\! & 3\!\\
0\! & 0\! & 0\!
\end{array}
\right)  \right]^{-1} \times
\nonumber\\
&  \times & \sum_{\mu=0}^{3}(-1)^{\mu}(2\!-\!\delta_{\mu0})\left(
\begin{array}
[c]{ccc}%
3 \!& k\! & 3\!\\
-\mu\! & 0\! & \mu\!
\end{array}
\right)\! e_{\mu}
\end{eqnarray}
The  algebraic equivalence of the two sets of intrinsic parameters
evidenced by above relation allows one to extend the literature
data of their values, independently of the way in which they have
been obtained.

\subsection{\label{f.contributions}Contributions to the CF potential}
The procedure outlined in subsection \ref{f.effe} leads to several
characteristic contributions to the CF potential $V$,
corresponding to different mechanisms. Their discussion in next
sections precedes a formulation of the enhanced angular overlap
model. Thus, it is advisable to separate out these which obey the
assumptions of the angular overlap model (the {\it AOM-consistent}
contributions), $V^{AOM}$, and the {\it residuum}, $V^{res}$:
\begin{equation}
V = V^{AOM} + V^{res}.
\end{equation}
In $V^{AOM}$ one can further distinguish the {\it primary}
($V^{pr}$) and {\it renormalization} ($V^{ren}$) components:
\begin{equation}\label{vaom}
V^{AOM} = V^{pr} + V^{ren}.
\end{equation}
$V^{pr}$ represents the Coulomb interaction (direct and exchange)
of the ligand electrons, and the potential of the nuclei. It
diverges from the {\it point charge model} (pcm)\cite{b29} due to
the {\it charge penetration} ($cp$)\cite{k52} and inter-ionic {\it
exchange} ($ex$)\cite{gm85a} effects included in $V^{pr}$:
\begin{equation}\label{pr}
V^{pr} = V^{pcm} + V^{cp}+V^{ex}+ V^{pr.sh}.
\end{equation}
The intra-ionic excitations on the metal ion induced by the
primary contribution lead to the AOM-consistent part of the {\it
shielding} potential $V^{pr.sh}$.

$V^{ren}$ comprises the main renormalization terms implied by the
ligand-metal charge transfer excitations - the {\it covalency}
contribution, $V^{co}$,\cite{n71} and non-orthogonality of the
free-ion wave-functions localized on adjacent ions: the {\it
overlap} contribution, $V^{ov}$,\cite{n71} and the {\it contact
shielding}, $V^{cs}$,\cite{gmf87}:
\begin{equation}\label{ren}
V^{ren} = V^{ov} + V^{cs} + V^{co}
\end{equation}
$V^{res}$ includes all the remaining terms. One can distinguish
$V^{nn.pol}$ - the contribution of electric multipoles induced on
ligands (polarization of nearest neighbors), $V^{fn}$  - the
electrostatic potential of the point charges and electric
multipoles induced on the all the ions outside the cluster
(electrostatic potential of further neighbors) and $V^{res.sh}$
which symbolizes the shielding correction to these electrostatic
potentials:
\begin{equation}\label{res}
V^{res} = V^{nn.pol}  + V^{fn}+ V^{res.sh} .%
\end{equation}

Due to the cancellation of various terms in the primary and
residual contributions, $V^{ren}$ or $V^{ov}$ in essence, turns
out to be the most important mechanism for the ionic compounds in
favor of AOM.\cite{n71,fgj86,gm92,nn00,mg00} This corollary is
supported also by the results presented in next section.

Explicit formulas for all the above contributions are given in
Appendix B.

\section{\label{angu}{\it Ab initio} calculations}
\subsection{\label{a.deta}Details of the calculations}
The compounds under consideration represent a variety of the
crystal structures. The point symmetry of the metal ion varies
from cubic one in UO$_2$ (CaF$_2$ structure, space group $Fm3m$ -
225), through $D_{2d}$ in UO$X$ ( PbFCl-type structure, $P4/nmm$ -
129),\cite{smwyf98} $C_{2v}$ in UTe$_2$ (orthorhombic space group
$Immm$ - 71),\cite{s96} up to $C_s$ in $\beta$-US$_2$ and
$\beta$-USe$_2$ (PbCl$_2$-type structure, $Pnma$ -
62).\cite{npts96}

Zero-order free-ion wave-functions have been generated using
self-consistent Dirac-Slater procedure ATOM\cite{kh77,koe} with
the stabilizing potential well determined by the Madelung
energy.\cite{w58} The calculations have been performed in the
crystallographic coordinate system. The lattice sums of static and
induced multipoles generated by the set of the crystal
electrostatic equilibrium equations (\ref{apbequi}) in Appendix
\ref{apb} have been calculated according to Eq. (\ref{apbbel})
using a modified version of the program CHLOE.\cite{fau,fg82} and
multipole polarizabilities from Ref. \onlinecite{swd79}. The
summation in the multipole expansion (\ref{apbbel}) was limited to
monopoles, dipoles and quadrupoles ($n = 0, 1, 2 $ ) in the
present calculations. The effect of the outer electrons occupying
the $6s$ and $6p$ closed shells of the metal ion have been
estimated via Sternheimer's shielding factors, $\sigma_k$, scaling
the corresponding $B_{kq}$'s\cite{sbp68,er83}. The charge
penetration, exchange and renormalization terms have been
calculated using the program LF developed for the $f$-electron
compounds.\cite{gaj}

The free-ion part of the effective Hamiltonian contains the
intraionic Coulomb repulsion of the $5f$ electrons of the U(4+)
ion, controlled by the Slater integrals $F^k$, $k = 2, 4, 6$,
spin-orbit coupling with the $\zeta_{5f}$ parameter and further
corrections of higher order (see Appendix \ref{apa}). The
"free-ion" parameters depend on the crystal, in which the metal
ion is embedded, and may vary in certain range, modifying the
inter-term spacing.\cite{cc85} Fortunately, these differences are
limited and they have rather negligible impact on the splitting of
the ground multiplet. Thus, we adopt the values obtained earlier
from the fitting of the optical spectra for U$^{4+}$ in
ThGeO$_{4}$ (in meV):\cite{gka97} $F^2 = 5339$, $F^4 = 4833$, $F^6
= 3018$, $\alpha = 3.72$, $\beta = -81.8$, $\gamma  = 148.8$, $M^0
= 0.124$, $M^2 = 0.069$, $M^4 =0.047$, $P^2 = 62.0$, $P^4 = 31.0$,
$P^6 = 6.2$, $\zeta_{5f} = 224.2$. The calculations of the
eigenvalues and eigenvectors and/or fitting of the measured
energies of electronic transitions have been performed with a set
of f-shell empirical computer programs developed by M. F.
Reid,\cite{rei} supplemented with a subprogram POD\cite{gaj} for
calculation of some thermodynamic quantities.
\subsection{\label{a.reli}Reliability of the results}
\begin{figure}
\includegraphics[width = 7.5cm]{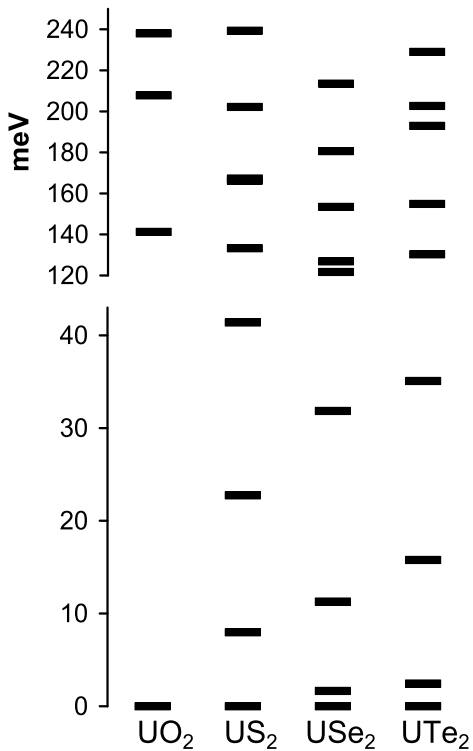}
\caption{\label{flev} Splitting of the uranium (4+) ground term
$^3H_4$ in the U$X_2$ crystals obtained by simultaneous
diagonalization of Hamiltonian (\ref{apaha}-\ref{apaha0}) with the
CF parameters determined from first principles. The lower part of
the diagram uses the scale enlarged several times as compared with
that for the upper part.}
\end{figure}

This type of ab initio calculations was performed previously for,
among the others, UO$_2$ and UOS ,\cite{glkm88,g00,g04} the
compounds for which comprehensive experimental data are available,
including optical  and/or inelastic neutron scattering (INS)
spectra.\cite{glkm88,otbhhhacbf88,abbcens95} Discrepancy of the
obtained theoretical $B_{kq}$ parameters and the phenomenological
values determined by a fitting to the experimental data did not
exceed 20-30\%. Now the calculations are extended to the U$X_2$
sub-series. For these compounds the published experimental data
are not so rich.\cite{npts96,stk95,nt96} We can reproduce certain
thermodynamic averages, e.g. the paramagnetic Van Vleck
susceptibility (see Appendix \ref{apc}) depicted in Fig.
\ref{fsus}. The corresponding energy levels are shown in Fig.
\ref{flev} and they are discussed later.
\begin{figure*}
\includegraphics[width = 17cm]{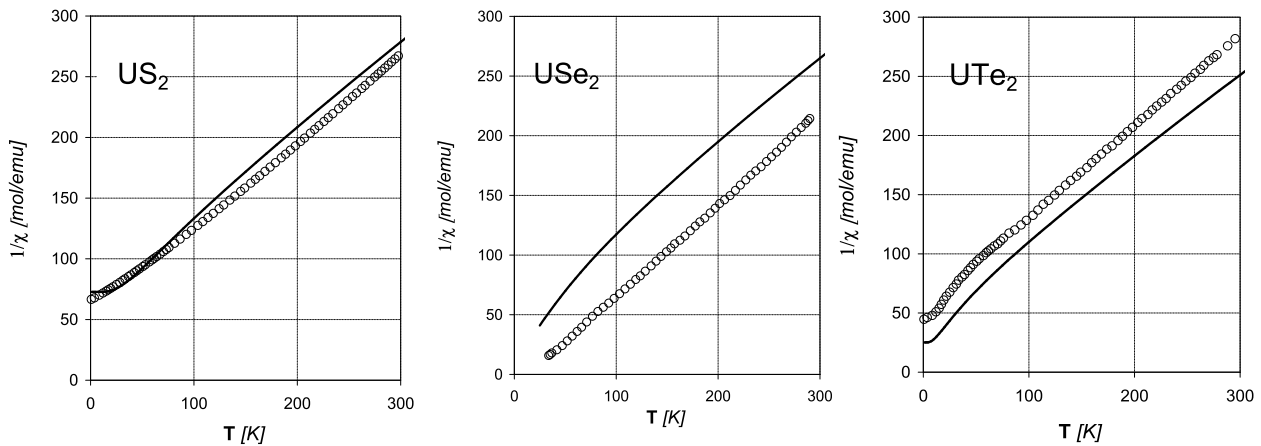}
\caption{\label{fsus} Reciprocal magnetic susceptibilities of
powdered dichalcogenides as a function of the temperature:
comparison of the experimental (open
circles)\cite{npts96,stk95,nt96} and theoretical (continuous
lines) curves obtained using the Van Vleck formula (\ref{apcvv}).}
\end{figure*}
The experimental slope, shape and the variation of the temperature
independent susceptibility at the lowest temperatures along the
series agree quite well with the results of the {\it ab initio}
calculations. A discrepancy between the theoretical and
experimental lines visible for USe$_2$ and UTe$_2$ can be
attributed to the interionic exchange not included in the model
calculations. Judging from the mutual shifts of the lines, the
interaction has to be of ferromagnetic type for USe$_2$ and
greater in the absolute value than the antiferromagnetic type
observed for UTe$_2$. The first excited level lying at 8.0, 1.6
and 2.5 meV for U$^{4+}$ ion in US$_2$, USe$_2$ and UTe$_2$,
respectively (see Fig. \ref{flev}) influences the magnetic
properties of these systems with a singlet ground state. Above 20
K, UTe$_2$ and especially USe$_2$ behave as if they had a
degenerate ground state. Relatively weak U-U exchange interaction
in USe$_2$ (U-U distance of 0.423 nm\cite{npts96}) turns out to be
sufficient to induce the long-range ferromagnetic order below 14
K.\cite{stk95} Note, that the first excited state has the lowest
energy just for this compound. USe$_2$ is the only dichalcogenide
that orders magnetically.\cite{npts96} No ordering in UTe$_2$ with
the shortest U-U distance 0.378 nm may be due to the fact that
only one of seven neighboring uranium ions is placed at that
distance, whereas the remaining four are placed at 0.490 nm and
two at 0.416 nm. A different curvature of the theoretical and
experimental reciprocal susceptibilities observed for USe$_2$
below 60 K may be ascribed to the magnetic fluctuations increasing
with temperature approaching the critical point, which were not
taken into account in the model calculations. The effective
magnetic moments, of 2.94 BM, 3.01 BM and 3.09 BM, are lower than
those derived from the experimental curves, of 3.25 BM 3.20 BM and
3.21 BM, reported for US$_2$, USe$_2$\cite{stk95} and
UTe$_2$,\cite{nt96} respectively.

Figure \ref{flev} shows the splitting of the ground term $^3H_4$
of the $U^{4+}(5f^2)$ ion in the U$X_2$ series, obtained by
simultaneous diagonalization of Hamiltonian
(\ref{apaha}-\ref{apaha0}), with the CF parameters determined from
first principles shown in Table \ref{tbkq} and the free-ion
parameters listed in Appendix \ref{apa}. These results have been
employed in calculations of the temperature dependencies of the
magnetic susceptibilities discussed above.
$B_{kq}$ parameters in Table \ref{tbkq} reveal no regularity along
the series that might be expected, for instance, from the
spectrochemical ordering of ligands. This observation is not only
a consequence of the inherent ambiguity of the CF parametrization
in low symmetry systems.\cite{rq04a} It would be difficult to find
any trend also for the energy levels shown in Fig.\ref{flev}.
Analogous behavior of the $An$O$X$ series has been ascribed to an
influence of the competing oxygen and chalcogenide groups in the
$W_{kq}^\mu$ coefficients (\ref{wkq}).\cite{g00,g04} Now it
becomes evident that also in the case of one type of anions the
coordination geometry may obscure the expected regularities so
clearly manifested, on the other hand, by the intrinsic AOM
parameters (cf. Table \ref{taom}). As we see in next section,
possible trends in CFP sets revealed elsewhere using an
independent method based on certain conserved quantities
associated with CF parameters\cite{rq04} may be governed by the
AOM-consistent part of crystal field.
\subsection{\label{a.aomc}AOM-consistent part of crystal field}
The results displayed in Table \ref{tbkq} are indicative of an
essential meaning of the AOM-consistent contributions.
Nevertheless they also show that the residual contribution may
become crucial for some parameters as in the case of $B_{44}$ for
US$_2$ or $B_{22}$ for UTe$_2$. More detailed data, which are
presented in Table \ref{tbkq} for USe$_2$, give some idea about
the role of the particular CF mechanisms. Mutual compensation of
the various primary components and importance of the
renormalization terms is clearly manifested there.
\begin{table*}
\caption{\label{tbkq} {\it AOM-consistent} (\ref{vaom}) and {\it
residual} (\ref{res}) contributions to the CF parameters
calculated from first principles for the U$X_2$ series in the
crystallographic coordinate systems.\cite{stand} The primary
contribution $V^{pr}$ (\ref{pr}), renormalization term $V^{ren}$
(\ref{ren}), ligand polarization $V^{nn.pol}$ (\ref{apbbel}) and
potential of the further neighbors $V^{fn}$ are specified for
USe$_2$ in parentheses. All values in meV.}
\begin{ruledtabular}
\begin{tabular}{l|ccc|ccccc|ccc}
&&US$_2$&&&&USe$_2$&&&&UTe$_2$& \\
&$V^{AOM}$&$V^{res}$&$V$&$V^{AOM}$&(= $V^{ren}$ + $V^{pr}$)&$V^{res}$&(= $V^{nn.pol}$ + $V^{fn}$)&$V$&$V^{AOM}$&$V^{res}$&$V$\\
\hline
$B_{20}$&65&-17&48&84&(73 + 11)&-10&(-24 +14)&74&38&13&51\\
$B_{21}$&-203&-3&-206&-200&(-194 - 5)&-6&(-1 - 5)&-206&&&\\
$B_{22}$&-79&-6&-85&-24&(-26 + 3)&-3&(-9 + 6)&-26&28&42&70\\
$B_{40}$&-88&2&-86&-71&(-52 - 19)&4&(4 - 0)&-67&-436&-184&-619\\
$B_{41}$&82&6&88&39&(32 + 7)&-5&(0 - 5)&34&&&\\
$B_{42}$&-274&-8&-282&-264&(-297 + 33)&-16&(6 - 22)&-280&364&79&442\\
$B_{43}$&257&45&301&186&(236 - 49)&26&(25 + 1)&212&&&\\
$B_{44}$&-60&-51&-111&-64&(-62 - 2)&-48&(-21 - 27)&-112&74&-26&48\\
$B_{60}$&174&-2&172&78&(80 - 3)&1&(2 - 1)&78&87&4&91\\
$B_{61}$&-233&-1&-233&-230&(-311 + 82)&-1&(-1 + 0)&-231&&&\\
$B_{62}$&394&5&398&340&(441 - 101)&3&(-5 + 9)&344&11&-6&6\\
$B_{63}$&171&-1&170&203&(248 - 45)&-1&(-1 - 0)&202&&&\\
$B_{64}$&37&18&55&34&(59 - 26)&9&(8 + 1)&43&-122&26&-96\\
$B_{65}$&-369&-10&-379&-320&(-383 + 63)&-4&(-4 - 0)&-324&&&\\
$B_{66}$&243&9&252&212&(262 - 50)&6&(2 + 4)&219&-151&0&-151\\
\end{tabular}
\end{ruledtabular}
\end{table*}
\begin{table}
\caption{\label{taom} Results of ab initio calculations of the AOM
parameters (in meV) corresponding to the {\it AOM-consistent} part
of the CF potential for various $An^{4+}-X^{2-}$ systems and the
average inter-ion distances $R_{av}$.}
\begin{ruledtabular}
\begin{tabular}{lcccc}
&$R_{av}$ [nm]&$e_{\sigma}$ & $e_{\pi}$ & $e_{\delta}$ \\
\hline
\multicolumn{2}{c}{U$^{4+}$-O$^{2-}$ in:}&&&\\
UO$_2$&0.237&350&209&64\\
UOS\footnotemark[2]&0.234&340&211&63\\
UOSe\footnotemark[2]&0.236&324&196&55\\
UOTe\footnotemark[2]&0.237&316&186&46\\
\multicolumn{2}{c}{U$^{4+}$-S$^{2-}$ in:}&&&\\
UOS&0.293&211&96&32\\
US$_2$&0.289&265&102&36\\
\multicolumn{2}{c}{U$^{4+}$-Se$^{2-}$ in:}&&&\\
UOSe&0.304&204&91&31\\
USe$_2$&0.301&240&88&30\\
\multicolumn{2}{c}{U$^{4+}$-Te$^{2-}$ in:}&&&\\
UOTe&0.325&185&90&31\\
UTe$_2$&0.317&190&88&29\\
\multicolumn{2}{c}{Np$^{4+}$-O$^{2-}$ in:}&&&\\
NpO$_2$&0.235&317&186&56\\
NpOS&0.233&304&181&53\\
NpOSe&0.235&291&169&47\\
\multicolumn{2}{c}{Np$^{4+}$-S$^{2-}$ in:}&&&\\
NpOS&0.291&191&87&30\\
\multicolumn{2}{c}{Np$^{4+}$-Se$^{2-}$ in:}&&&\\
NpOSe&0.302&183&82&29\\
\end{tabular}
\end{ruledtabular}
\footnotetext[2]{From Ref.~\onlinecite{g00}.}
\end{table}
Handling each non-equivalent $ML$ pair independently in compounds
like US$_2$ (with six different $ML_{t}$ distances and two
non-equivalent crystallographic positions of the sulphur ion)
would multiply the number of the intrinsic parameters, making the
model practically intractable. Usually, the dependence of the
intrinsic parameters $e_{\mu}$ (or their renormalized counterparts
$s_{\mu}$ (\ref{e2e})) on the $ML$ distance, $R$, is assumed to
have a definite character\cite{nn00,g00}. In a limited range of
distances around an average value for a given $ML$ bond, it is
approximated by the power function (\ref{r2r}). The power
exponents, $\alpha_{\mu}$, are treated then as an additional
characteristic of the $ML$ bond that allows one to maintain the
minimal number of independent parameters.

The simulations of the $e_{\mu}(R)$ functions have been performed
for all the uranium linear ligators occurring in the series
U$^{4+}-X^{2-}$. As Eqs. (\ref{apbov}-\ref{apbdel}) in Appendix
\ref{apb} show, the functions are determined by squares of
metal-ligand overlap integrals and Madelung energies of the ions
in crystals. To ensure comparability of the results obtained for
different $ML$ pairs, a virtual $ML_2$ crystal of the CaF$_2$
structure has been employed, where ligands form a cube with its
center occupied by the metal ion. The Madelung potential at the
$L$ and $M$ sites in this structure is given by the formulae: $U^L
= -8.14e/a$ and $U^M = 15.13e/a$, respectively, whereas the
lattice constant $a$ is related to the $ML$ distance $R$ by the
expression $R = \sqrt{3}a/4$. The intrinsic parameters depend on
$U^L$ and $U^M$ in a non-trivial way through the zero-order
wave-functions and explicitly due to the renormalization terms
(\ref{apbov}, \ref{apbcov}). This implies the use of different
free-ion wave-functions for each $R$. The calculations have been
performed for seven values of $R$ distributed uniformly between
0.18 nm and 0.40 nm. The radius $D$ of the stabilizing potential
well for the negative ions of the $U^L$ depth has been assumed to
be equal to $-2e/U^L$.

\begin{figure}
\includegraphics[width = 8.5cm]{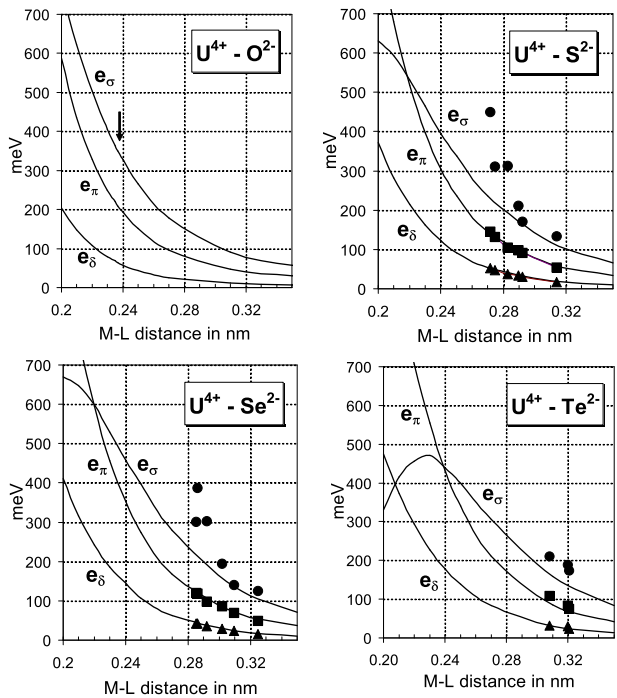}
\caption{\label{fdis} Dependence of the inter-ion effective
interaction parameters of the angular overlap model on
metal-ligand distance for the U$^{4+} - X^{2-}$ pairs in a
hypothetical crystal of CaF$_2$ structure. The circles, squares
and triangles represent actual intrinsic parameters $e_\sigma$,
$e_\pi$ and $e_\delta$ obtained for U$X_2$.}
\end{figure}

\begin{figure*}
\includegraphics[width = 16cm]{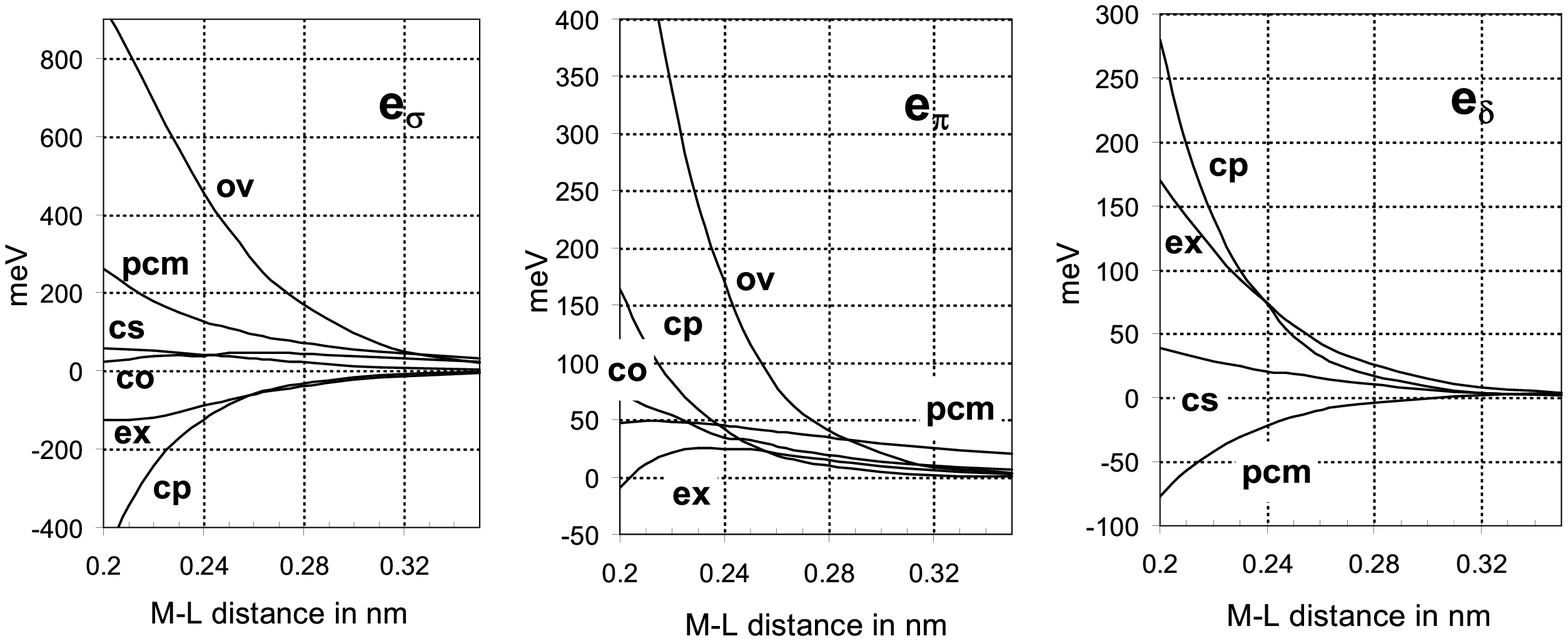}
\caption{\label{fcom} Distance dependence of the main
contributions to the $e_{\mu}$ parameters: point charges ($pcm$),
overlap ($ov$), covalency ($co$), charge penetration ($cp$),
exchange ($ex$), contact shielding ($cs$).}
\end{figure*}

The results are shown in Fig. \ref{fdis}. Due to the predominating
renormalization terms (cf. Fig. \ref{fcom}), the $e_{\mu} (R)$
functions have an exponential character for the $ML$ distances
around the average values. For larger distances, where the
electrostatic contributions dominate, they take a form of a
polynomial. The slope of $e_{\sigma}(R)$ decreases in the limit of
short distances, more visibly on going from the oxide to
telluride. The function $e_{\sigma}(R)$ reaches even a maximum for
the U$^{4+}$-Te$^{2-}$ bond. Fig. \ref{fcom} indicates that the
importance of the charge penetration rapidly increases as compared
with that for other mechanisms. Since this contribution has
opposite sing for $e_{\sigma}$ and $e_{\pi}$, the ratio
$e_{\pi}/e_{\sigma}$ increases in the limit of small distances.%

A similar behavior, although not so evident, can be deduced from
the data presented in Table II of Ref. \onlinecite{sh94} for the
Pr$^{3+}$-Cl$^{-}$, if the Coulomb contributions are scaled using
the Sternheimer shielding factors and the SPM parameters are
converted into the AOM ones by Eq. (\ref{e2b}). An open question
is whether the observed increase of the $e_{\pi} /e_{\sigma}$
ratio at the lowest distances characteristic for dense systems is
a true property of metals. We may only note that analogous
distance dependencies of the AOM parameters for the
Sm$^{2+}$-Cl$^{-}$ and Sm$^{2+}$-F$^{-}$ systems derived from the
data reported in Ref. \onlinecite{sb98} did not confirm this
conjecture.

The model calculations taking into account the variation of
Madelung energy with the $ML$ distance lead to lower values of the
power exponents $\alpha_{\mu}$ as compared with earlier
estimations.\cite{g00} The AOM parameters determined for the
actual crystals $An$O$X$ and $AnX_2$ are listed in Table
\ref{taom} and shown in Fig. \ref{fdis}. Each $ML$ distance
occurring in these compounds is represented in Fig. \ref{fdis} by
three points corresponding to the $e_{\sigma}$, $e_{\pi}$, and
$e_{\delta}$ parameters. We see that they do not follow the smooth
lines discussed above. The crucial $e_\sigma$ parameter turns out
to be simultaneously the most irregular one. This is because it
reflects the discontinuity of the Madelung potential most. In
light of the present calculations, the properties (i-v) of the AOM
parameters specified in section \ref{f.angu} should be treated
with caution not only because of very existence of the residual
contributions but also because of the revealed here inherent
irregularity of the AOM-consistent contributions.
\section{\label{enha}Enhanced angular overlap model}
\subsection{\label{e.form} Formulation}
In view of apparent irregularity of the AOM parameters and the
erratic behavior of the residual contribution (\ref{res})
discussed in the preceding section and manifested in Fig.
\ref{fdis} and Table \ref{tbkq} we propose to link the
phenomenological AOM approach with the {\it ab initio}
calculations in the enhanced angular overlap model (EAOM). Our
model assumes each ordinary CF parameter to be composed of two
components: the main, adjustable one, parametrized according to
Eqs. (\ref{bkqem}-\ref{e2e}) and the fixed residuum,
$B_{kq}^{res}$ corresponding to the potential (\ref{res}),
determined from Eq. (\ref{apbbel})):
\begin{equation}\label{eaom2bkq}%
B_{kq}=\sum_{\mu}W_{kq}^{\mu}e_{\mu}+B_{kq}^{res}
\end{equation}
with $e_\mu$  playing the role of the phenomenological EAOM
parameters. The separation of the residual, off-AOM contribution
from the global CF parameters supports a more precise and
consistent AOM parametrization of the remainder. In consequence,
the intrinsic character of the corresponding EAOM parameters is
genuinely maintained. EAOM is not much more complicated than the
parental AOM but the main inherent shortcoming of the latter is
removed by excluding explicitly the off-AOM contributions from the
simplified phenomenological treatment. These off-AOM contributions
are estimated in our model from the first principles. Note, that
very construction makes the model exact (at least in the frames of
the one-electron approximation) provided the off-AOM contributions
are determined  precisely, what, of course, is hardly possible.

The AOM-consistent contributions listed in Table \ref{taom} can be
regarded as a crude theoretical estimation of the EAOM parameters.
Their more accurate determination requires an involved
self-consistent approach to account for the energy-dependent
parameter-renormalization terms.\cite{ghw81} Thus, treating this
part of the CF potential as a phenomenological quantity allows one
to handle these self-energy effects in the simplest possible way.

Naturally, the EAOM parameters obtained from the fitting of the
observed electronic energy levels cannot have the same values as
the AOM parameters derrived from the analogous fitting within the
conventional AOM approach. Moreover, the rules (i-v) observed for
the usual AOM parameters (see section \ref{f.angu}) seem more
justifiable in the case of the theoretical, AOM-consistent
contributions and thus, also for the EAOM parameters.
Nevertheless, in light of the calculations presented in section
\ref{a.aomc}, still they remain acceptable in a rather limited
range of the $ML$ distances, in vicinity of their averaged values.

Compounds containing several groups of symmetry-equivalent ligands
require the $s_{\mu}^t$ ratios in the geometrical coefficients
$W_{kq}^{\mu}$ (\ref{wkq}) to be determined. The observed above
(see {section \ref{a.aomc} and Fig. \ref{wkq}) irregularity of the
EAOM parameters due to, among the others, the Madelung energy, can
be reproduced in the model by employing the {\it ab initio} values
of these ratios. Thus, it is advisable to calculate $s_{\mu}^t$'s
for each individual $ML_t$ pair rather than applying the
approximation (\ref{r2r}).

The same theoretical calculations may also be employed to
formulate further, more restrictive parametrizations, which may be
helpful in the case of the most complicated experimental data. For
instance, fixing both the ratios
\begin{equation}\label{e2es}%
s_{\mu\sigma}=e_{\mu}/e_{\sigma}, \quad \mu=\pi, \delta
\end{equation}
equal to their theoretical values leads to a single-parameter
version of the model:
\begin{equation}\label{eaom2bkq0}%
B_{kq}=B_{kq}^{res}+\widetilde{W}_{kq}\tilde{e}
\end{equation}
with
%
\begin{eqnarray}\label{wkq0}
\widetilde{W}_{kq}=\frac{2k+1}{7}\left[  \left(
\begin{array}
[c]{ccc}%
3 \!& k\! & 3\!\\
0\! & 0\! & 0\!
\end{array}
\right)  \right]^{-1}\sum_{\mu}(-1)^{\mu}(2\!-\!\delta_{\mu0})\times\nonumber\\
\times\left(
\begin{array}
[c]{ccc}%
3 \!& k\! & 3\!\\
-\mu\! & 0\! & \mu\!
\end{array}
\right)\sum_t C_{q}^{(k)\ast}(\Theta_{t},\Phi_{t}) s_{\mu}^t s_{\mu\sigma}%
\end{eqnarray}
%
The $\tilde{e}$ parameter in Eq. (\ref{eaom2bkq0}) corresponds
directly to $e_{\sigma}$  but also the $e_{\pi}$ and $e_{\delta}$
contributions enter into the model through the $s_{\mu\sigma}$
ratios in the above $\widetilde{W}_{kq}$ coefficients. Hence, a
variation of $\tilde{e}$ modifies also the $e_{\pi}$ and
$e_{\delta}$ contributions to the crystal field in proportions
given by $s_{\mu\sigma}$. Note, that Eq. (\ref{eaom2bkq0}) without
the residual part would be a simple scaling of the crystal field
effect predicted by the AOM-consistent part of the {\it ab initio}
calculations.

An intermediate, two-parameter version of the model can be defined
in several ways by fixing any of the $s_{\mu\sigma}$ ratios or
their combinations. These single- and two-parameter versions,
should not be confused with models commonly employed for the
transition elements\cite{ghw81,nn00} which omit simply the $\pi$
and $\delta$ contributions or only the $\delta$ contribution.

Generalization of the model and the equations
(\ref{eaom2bkq}-\ref{wkq0}) to systems containing different
ligands is straightforward (see the example given below).

\subsection{\label{e.appl} Applications}
Due to the highly reduced number of the free parameters (up to the
single-parameter version, Eq. (\ref{eaom2bkq0})), the simplified
EAOM parametrizations of the CF effect may be especially helpful
in all these cases of incomplete or more complicated experimental
data so frequently met in the most interesting $f$-electron
systems. Three possible types of applications of EAOM are
exemplified in what follows.

The first one, already discussed in Ref. \onlinecite{g04},
concerns interpretation of certain magnetic properties of NpO$X$
where $X$ = O, S and Se. The procedure might be seen as a general
method of interpretation of the electronic properties based on
transferability of the intrinsic parameters between different
compounds, using the phenomenological CF parameters for the
specimens assumed to be known, i.e. the compounds, the electronic
structure of which is believed to be reliably determined. In the
example under consideration, the parameters $(B_{kq})^{U}$
reported for UO$_2$, UO$X$, the ones of the most widely
investigated uranium compounds, were employed.\cite{glkm88,g00}
The corresponding AOM parameters $e_{\mu}^{UO}$ and $e_{\mu}^{UX}$
were determined from system of equations (\ref{eaom2bkq}) and its
version adapted to two different ligands, oxygen and chalcogenide,
\begin{equation}\label{eaom2bkq1}%
B_{kq}=B_{kq}^{res}+\sum_{\mu}\left(W_{kq}^{\mu,O}e_{\mu}^O+W_{kq}^{\mu,X}e_{\mu}^X\right)
\end{equation}
exploiting the {\it ab initio} values of the $s_{\mu}^t$ ratios.
In the second phase, the procedure was reversed to estimate
unknown CFP's for the less explored experimentally neptunium
oxychalcogenides. First, the $e_{\mu}^{NpO}$ and $e_{\mu}^{NpX}$
parameters were derived by scaling their uranium counterparts in
terms of the ratios of the corresponding theoretical values:
$(e_{\mu}^{NpX}/e_{\mu}^{UX})_{calc}$. The $B_{kq}^{Np}$
parameters obtained, again, from Eqs. (\ref{eaom2bkq},
\ref{eaom2bkq1}) were, in the final step, adjusted to match the
experimental values of the ordered magnetic moments. In this
phase, a simultaneous diagonalization included, apart from the CF
Hamiltonian and the "free-ion" interactions, the intermetallic
exchange interaction in the zero-temperature mean-field
approximation. The magnetic properties of the neptunium
oxychalcogenides, including the anisotropic ground state magnetic
moment, the temperature dependencies of the paramagnetic
susceptibility and the magnetization at 0 K, were described
satisfactorily in this approach (see Ref. \onlinecite{g04} for
further details). The example illustrates efficiency of the method
based on EAOM in interpretation of the electronic structure in the
case of inconclusive experimental data. The intrinsic parameters
collected in Table \ref{taom}, may serve as a tentative source of
the relations between them in further applications of EAOM.

The second example concerns interpretation of the thermodynamic
and magnetic properties\cite{abchmt84,t87} and inelastic neutron
scattering (INS) spectra of UOS.\cite{abflo89,abbcens95} In the
analysis of the experimental data,\cite{g00} the initial energy
levels assignment was deduced on the grounds of first-principles
calculations. The fitting of the INS data\cite{abflo89,abbcens95}
was performed in two steps. First, the AOM was applied to obtain
the identified INS transition energies. Subsequent refinement of
the corresponding usual CF parameters yielded agreement with the
remaining experimental data.\cite{abchmt84,t87,abbcens95} A
special attention was paid to the ordered magnetic moment
$\mu_{ord}$ and relative intensities of the INS transitions.
However, the problem with the infinite number of acceptable
solutions could not be resolved. Namely, the experimental data for
this tetragonal $C_{4v}$ system could be satisfactory reproduced
with any value of the $B_{20}$ in the range of $-50$ to $-223$
meV, provided the remaining parameters, $B_{40}$, $B_{44}$,
$B_{60}$, $B_{64}$ were simultaneously adjusted to the given
$B_{20}$. Some features of the electronic structure were varying
with the CF parameters in this ambiguity range but they were not
detectable in $\mu_{ord}$ and in the position and shape of the
most intense INS lines below 100 meV. In addition, it was
demonstrated\cite{g00} that the INS excitations to two of the
highest levels of the $^3H_4$ term, $\Gamma^2$  and
$\Gamma^{1(2)}$, could be invisible if $B_{20}$ had decreased
below -146 meV under actual incident neutron energy and angle,
just as it was observed. Eventually, the set of $B_{kq}$
parameters corresponding to the threshold value of $B_{20}=-146$
meV was accepted as the closest one to the initial estimation.

The above problem disappears if EAOM is applied with its inherent
constrains. To start with, the single-parameter version of EAOM,
Eq. (16), is adapted to account for the two different anions
occurring in the coordination sphere - the oxygen and sulphur:
\begin{equation}\label{eaom2bkq00}%
B_{kq}=B_{kq}^{res}+\widetilde{W}_{kq}^O\tilde{e}^O+\widetilde{W}_{kq}^S\tilde{e}^S
\end{equation}
with coefficients $W$ given by Eq. (\ref{wkq0}), $B_{kq}^{res}$
and the remaining data taken from Ref. \onlinecite{g00}. It turns
out that this simplest version of EAOM, with only two effective
parameters $\tilde{e}^O$ and $\tilde{e}^S$, describes fairly well
the three electronic INS transitions:\cite{abflo89,abbcens95} 74
meV ($\Gamma^5(1)\rightarrow\Gamma^3$), 83 meV
($\Gamma^5(1)\rightarrow\Gamma^4$), and 87 meV
($\Gamma^5(1)\rightarrow\Gamma^5(2)$). These intervals can be
easily matched by a minor refining of the {\it ab inito} ratios
$s_{\pi\sigma}$. Figure \ref{fins} (c) shows the resulting
energies and the transition probabilities estimated in terms of
the squares of the matrix elements of the Zeeman operator between
the initial and final states.\cite{b72} The Debye-Waller and
$5f$-electron form factors are not taken into account since we are
interested only in comparison of various numerical simulations
without a direct reference to the experimental recordings. The
Gaussian shape of the figured transitions have only an
illustrative character. The present simulation of the INS
transitions does not differ much from the previous one shown in
Fig. \ref{fins} (b).\cite{g00} Taking into account a finite width
of the experimental lines and limited range of the measurable
energies both the simulations seem acceptable. In particular, the
main lines around 80 meV are quite similar to the measured
ones.\cite{abflo89,abbcens95} The line at about 40 meV was neither
observed nor excluded by the experimental data since it is
relatively weak and lies in the region of the phonon sidebands.
The ordered magnetic moment of 2.11 MB as determined from the
obtained wave-functions, is higher than the observed one of 2.00
MB.\cite{t87} Nevertheless it can be reduced to about 2.04 BM due
to the overlap and covalency effects.\cite{gm85} Our
phenomenological results can be compared with {\it ab inito}
predictions. The parameter $\tilde{e}^O = 370.2$ meV is about 10\%
higher than the theoretical value of $e_{\sigma}$ from Table
\ref{taom}, whereas $\tilde{e}^S = 157.1$ meV is lower than its
counterpart. The differences between the corresponding
phenomenological and theoretical parameters become even more
pronounced in the case of the $B_{kq}$ parameters listed in Table
\ref{tuos}. Taking into account inherent limitations of the {\it
ab inito} calculations due to numerous approximations and
uncertainty of the Sternheimer shielding factors, dipole and
quadrupole polarizabilities, such a divergence seems inevitable.
We assume the ratios of the theoretical values of the EAOM
parameters, (\ref{e2e}, \ref{e2es}), used in formulation of the
approximate phenomenological models, Eqs. (\ref{eaom2bkq},
\ref{eaom2bkq0}), to be more reliable than the theoretical
parameters themselves because of an expected partial cancellation
of the calculation errors. The present EAOM values of the $B_{kq}$
parameters lie between those obtained using the conventional AOM
and the refined model from Ref. \onlinecite{g00}. The divergence
between the sets of the phenomenological parameters, especially
between these in the two last lines of Table \ref{tuos}, is
moderate. Further INS investigations with higher incident neutron
energies could decisively verify the predicted positions of the
$\Gamma^2$ and $\Gamma^{1(2)}$ levels.

\begin{figure}
\includegraphics[width = 7.5cm]{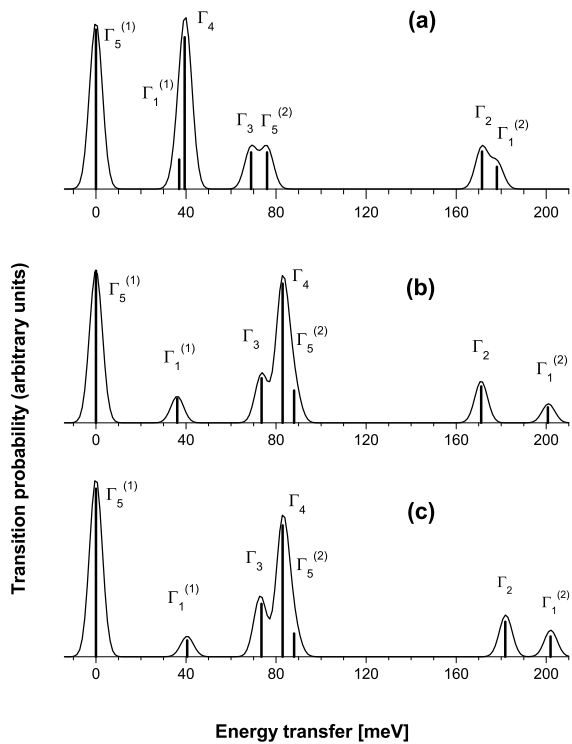}
\caption{\label{fins} Simulation of the INS spectra of UOS in
various crystal field models: (a) - {\it ab initio} calculations,
(b) - refined model from Ref. {\onlinecite{g00}}, (c) - EAOM.}
\end{figure}
\begin{table}
\caption{\label{tuos} Comparison of various {$B_{kq}$} sets
obtained for UOS (in meV).}
\begin{ruledtabular}
\begin{tabular}{lccccc}
& $B_{20}$ & $B_{40}$ & $B_{44}$&$B_{60}$ & $B_{64}$\\
\hline
{\it Ab initio} calculations &-180&-545&-119&424&104\\
AOM&-117&-567&-11&632&250\\
Model from Ref.\onlinecite{g00} &-147&-579&-39&595&447\\
EAOM &-159&-624&-21&627&380\\
\end{tabular}
\end{ruledtabular}
\end{table}

The low symmetry crystals, U$X_2$, considered in section
\ref{a.reli} may serve as a next example of application of EAOM.
The model allows one to reduce 15 (US$_2$, USe$_2$) or 9 (UTe$_2$)
$B_{kq}$ parameters specific for the point group symmetries in
these compounds to only 1-3 EAOM parameters. Note, that very
construction of any version of EAOM, including the
single-parameter one, Eq. (\ref{eaom2bkq0}), ensures accuracy of
the electronic structure simulation not worse than that obtained
from the {\it ab initio} calculations. This is because the {\it ab
initio} calculations determine not only the starting values of the
parameters but also the coefficients $W$ and residual off-AOM
terms in Eqs. (\ref{eaom2bkq},\ref{eaom2bkq0}). Thus, with EAOM
one can try to reproduce the electronic structure of any compound
even if the available experimental data are limited merely to a
single thermodynamic characteristic like the magnetic
susceptibility discussed in section \ref{a.reli}. In the specific
case of U$X_2$, the theoretical curves displayed in Fig.
\ref{fsus} represent simultaneously the initial phenomenological
dependencies in our model. The corresponding EAOM parameters
listed in Table \ref{taom} are thus the natural and right starting
values in subsequent steps of the further interpretation of the
electronic structure.

\section{\label{conc}Concluding remarks}
The crystal field potential in the quasi-phenomenological enhanced
angular overlap model (EAOM) proposed here is divided into two
parts: the main part, adjustable, which comprises the
contributions amenable to the conventional AOM parametrization and
the residual one, fixed, representing all the remaining terms. The
fixed, off-AOM residual part contains the electrostatic
contributions of further neighbors and the ligand polarization
contribution, accuracy of determination of which hinge on the
available ionic polarizabilities and the shielding factors. Much
more complicated renormalization terms in the CF potential are
handled in the phenomenological way together with the core, charge
penetration and interionic exchange contributions. The
parametrization applied is known from the usual AOM approach. Its
simplicity is a consequence of the axial local symmetry of the
encoded contributions and their additivity.

Aptness of such a discrimination between the contributions is
illustrated by an increasing amount of examples of first
principles calculations in the literature and also by the results
presented in this paper. Additionally, the calculations provide
the ratios of the intrinsic parameters which may be employed in
further applications of EAOM. The simulations performed for the
virtual U$X_{2}$ crystal of CaF$_{2}$ symmetry give an idea about
dependence of the EAOM parameters on metal-ligand distance. They
show the evolution of the mutual relations between the
AOM-consistent contributions. The observed behavior of the
AOM-consistent contributions to the intrinsic parameters in the
limit of the shortest metal-ligand distances touches a more
complex problem of the CF effect in metallic systems. A comparison
of these simulation with the {\it ab initio} calculations for the
actual U$X_{2}$ compounds evidences certain volatility of the
intrinsic parameters. It turns out that the assumption, commonly
accepted for models based on the superposition anzatz
(conventional AOM or SPM and their various modifications), saying
that the intrinsic parameters are smooth functions of metal-ligand
distance is not always valid due to various Madelung energies of
the ions in the real crystals and a high sensitivity of the
intrinsic parameters to these energies. This points to necessity
of enforcing each particular implementation of the
phenomenological model with the {\it ab initio} calculations for
an actual specimen just as it is postulated in EAOM.

EAOM ensures a highly compact description of the electronic
structure of even so complex ionic systems as the actinide
crystals. The examples discussed in the paper represent various
kinds of conceivable applications. The case of experimental data
limited to only certain magnetic characteristics of systems of low
crystal symmetries is illustrated for the U$X_{2}$ series. In
other example, the electronic structure for the NpO$X$ is
predicted by transforming the intrinsic parameters for the
corresponding UO$X$ compounds and exploiting relations between the
parameters calculated {\it ab initio}. Various experimental data
are available in the case of UOS. They include the magnetic and
thermodynamic properties as well as the inelastic neutron
scattering spectra. Nevertheless, the electronic structure cannot
be resolved within the conventional CF parametrization scheme
without additional assumptions of rather heuristic nature. EAOM is
shown to be capable of describing of all of the observed
properties unequivocally.

EAOM is open for further improvements. We mention here about a
more accurate potential that could be obtained from the lattice
self-consistent calculations\cite{ghw81,mg00} to correct the
zero-order wave-functions used in the calculations. Direct
evaluation of multipole polarizabilities and shielding factors for
a given specimen may be important for calculations of the residual
part of the crystal field, since these quantities depend on the
Madelung energies of the ions.\cite{swd79} Perhaps, an analogous
idea of partitioning of the CF potential into phenomenological and
fixed parts may be also applied to much more complicated metallic
systems.

\begin{acknowledgments}
The author thanks prof. Robert Tro{\'c} (ILT\&SR, Wroc{\l}aw) for
providing the source recordings of the susceptibility data for
U$X_2$, dr. Michael F. Reid (University of Canterbury,
Christchurch, New Zealand) for the f-shell programs, dr. Michelle
Faucher (Clamart, France) for the CHLOE program and prof. Jacek
Mulak (ILT\&SR, Wroc{\l}aw) for careful and critical reading of
the manuscript, .
\end{acknowledgments}

\appendix
\section{\label{apa}}
The parametric Hamiltonian\cite{w65,liu,mg00} contains the
free-ion, spherically symmetric part, $H_0$, and crystal-field
potential, $V$,
\begin{eqnarray}\label{apaha}
H = H_0 + \sum_{i}V({\bf r}_{i}/r_{i}),
\end{eqnarray}
where  the summation index $i$ runs over all $f$-electrons. The
free-ion part can be written as follows:
\begin{eqnarray}\label{apaha0}
H_0 = &&E_{ave}+\sum_{\text{k=2,4,6}}F^{\text{k}}(nf,nf)f_k +
\zeta_{5f}  A_{S0} + \nonumber\\
&&\alpha L (L+1) + \beta G (G_2)+ \gamma G(R_7)+ \nonumber\\
&&\sum_{\text{k=0,2,4}}M^j m_j + \sum_{\text{k=2,4,6}}P^k p_k
\end{eqnarray}
where $E_{ave}$ is the spherically symmetric one-electron part of
the Hamiltonian, $F^k(nf, nf)$ and  $\zeta_{5f}$ represent the
radial integrals due to the electrostatic and spin-orbit
interactions, while $f_k$ and $A_{SO}$ are the angular operators
corresponding to these interactions, respectively. The $\alpha$,
$\beta$ and $\gamma$ parameters are associated with the two-body
correction terms. $G$(G$_2$) and $G$(R$_7$) are Casimir operators
for the G$_2$ and R$_7$ groups and $L$ is the total orbital
angular momentum. The electrostatically correlated spin-orbit
perturbation is represented by the $P_k$ parameters and those of
the spin-spin and spin-other-orbit relativistic corrections by the
$M_j$ parameters. The operators associated with these parameters
are designated by $m_j$  and $p_k$  respectively.

\section{\label{apb}}
Main contributions to the one-electron CF potential in
non-metallic crystals are listed here. For derivations and a
detailed discussion see Ref. \onlinecite{mg00}. Some numerical
questions related to multicenter integrals and summation of weakly
convergent infinite series are dealt with in Refs.
\onlinecite{gm85a, gmf87, fg82}.

From definition, the AOM-consistent part of the CF potential and
thus the AOM-consistent contributions in Eqs. (\ref{pr}-\ref{ren})
are partitioned due to ligands according to Eq. (\ref{sup}).
Therefore, without lose of generality we can consider only single
ligand potential $v_{t}$ in the local coordinates centered at the
metal site, the z axis of which is directed towards the ligand.
Due to the axial symmetry of the ligand potential the orientation
of the x and y axes is immaterial. Atomic units are applied
throughout this section.
        \begin{equation}\label{apbpcm}
        v^{pcm}_t({\bf r})=
        \frac{2}{|{\bf r}-{\bf R_t}|}
        \end{equation}
        \begin{equation}\label{apbcp}
        v^{cp}_t({\bf r})=
        \sum_{\tau}\left[\hat{J}(\chi_{t\tau}({\bf r}),\chi_{t\tau}({\bf r}))-
        \frac{8}{|{\bf r}-{\bf R_t}|}\right]
        \end{equation}
        \begin{equation}\label{apbex}
        v^{ex}_t({\bf r})=
        -\sum_{\tau}\hat{K}(\chi_{t\tau}({\bf r}),\chi_{t\tau}({\bf r}))\
        \end{equation}
        \begin{eqnarray}\label{apbov}
        v^{ov}_t & = &
        \sum_{\tau}
        |\chi_{t\tau}\rangle\langle\chi_{t\tau}|\biglb[
        \langle \varphi|\hat{h}_{0}|\varphi\rangle - 2\hat{h}_{0} + 
        \nonumber \\
        & + &
        \sum_{t^{\prime},\tau^{\prime}}
        (\hat{h}_{0}-\hat{J}(\varphi,\varphi))
        |\chi_{t^{\prime}\tau^{\prime}}\rangle
        \langle\chi_{t^{\prime}\tau^{\prime}}|\bigrb]
        \end{eqnarray}
        \begin{eqnarray}\label{apbcs}
        v^{cs}_t & = &
        \sum_{\nu}\sum_{\tau}
        \biglb{\{}
        |\langle \xi_{\nu}|\chi_{t\tau}\rangle|^{2}
        [2\hat{J}(\xi_{\nu},\xi_{\nu})-\hat{K}(\xi_{\nu},\xi_{\nu})] -
        \nonumber \\
        & - &
        2\langle \xi_{\nu}|\chi_{t\tau}\rangle
        [2\hat{J}(\chi_{t\tau},\xi_{\nu})-\hat{K}(\chi_{t\tau},\xi_{\nu})] +
        \nonumber \\
        & + &
        \sum_{\stackrel{\scriptstyle t^{\prime},\tau^{\prime}}{
        t^{\prime}\neq t}}
\langle \xi_{\nu}|\chi_{t\tau}\rangle \langle
\chi_{t^{\prime}\tau^{\prime}}|\xi_{\nu}\rangle  \times
                \nonumber \\
        & \times &
        [2\hat{J}(\chi_{t\tau},\chi_{t^{\prime}\tau^{\prime}})-
          \hat{K}(\chi_{t\tau},\chi_{t^{\prime}\tau^{\prime}})]
        \bigrb{\}}
        \end{eqnarray}
        \begin{equation}\label{apbcov}
        v^{cov}_t=
        \sum_{\tau}\frac{\tilde{h}_{t\tau}|\chi_{t\tau}\rangle
        \langle\chi_{t\tau}|\tilde{h}_{t\tau}}{\Delta_{t\tau}}
        \end{equation}
$\varphi_m$ and $\xi_{\nu}$ in above equations denote the $5f$ and
$6s$, $6p$ orbitals of the metal ion obtained from the
Dirac-Slater calculations for the free ion, $\chi_{t\tau}$ stand
for the analogous $ns$ and $np$ orbitals of the anion and $m$,
$\nu$, $\tau$ are the corresponding sets of quantum numbers.
$\hat{J}$ and $\hat{K}$ are the usual Coulomb direct and exchange
operators (see for instance Ref. \onlinecite{mg00}),
       \begin{equation}\label{apbhion}
 \hat{h}_{0}= \frac{-\nabla^2}{2} + v^M_0+\sum_t v^L_{0t} +V^{fn}  \text{  ,}
        \end{equation}
where $v^M_0$ and $v^L_{0t}$ are the Dirac-Slater free-ion metal
and ligand potentials.
\begin{equation}\label{apbhtyl}
\tilde{h}_{t\tau} = h_{t\tau} - \sum_{t^{\prime}\tau^{\prime}}
\sum_{{t^{\prime}}^{\prime},{\tau^{\prime}}^{\prime}}
|\chi_{t^{\prime}\tau^{\prime}}\rangle
\langle\chi_{t^{\prime}\tau^{\prime}}|h_{t\tau}|
\chi_{{t^{\prime}}^{\prime}{\tau^{\prime}}^{\prime}}
\rangle\langle\chi_{{t^{\prime}}^{\prime}}|
\end{equation}
\begin{equation}\label{apbhtt}
h_{t\tau} = h_{0} - \hat{J}(\chi_{t\tau},\chi_{t\tau})
\end{equation}

        \begin{equation}\label{apbdel}
        \Delta_{t\tau}=
        \langle\varphi|h_{t\tau}|\varphi\rangle-
        \langle\chi_{t\tau}|h_0|\chi_{t\tau}\rangle
        \end{equation}
$v^{pcm}_t({\bf r})$ represents only ligands as point charges
(monopoles).

The remaining contributions are convenient to be presented as a
part of the electrostatic model, which includes all ions in the
crystal represented by a sequence of point monopole, dipole,
quadrupole etc.:
\begin{equation}\label{apbvel}
V^{el} = V^{nn.pol} + V^{fn} + V^{pcm}.%
\end{equation}
Note, that the two first terms on righthanded side of Eq.
(\ref{apbvel}) are part of $V^{res}$ in Eq. (\ref{res}), whereas
the third term enters into $V^{pr}$ in Eq (\ref{pr}). The
electrostatic potential $V^{el}$ of point multipoles can be
expanded into series of spherical harmonics, similarly as the
global CF potential in Eq. (\ref{1}). The corresponding
contribution to the $B_{kq}$ parameters have the following
form:\cite{mg00}
\begin{widetext}
        \begin{eqnarray}\label{apbbel}
        B_{kq}^{el}
        & = &
        \langle r^{k}\rangle
        \sum_{t}\sum_{n}\sum_{\mu}(-1)^{k+q+\mu}
        \left[
        \frac{(2k+2n+1)!}{2^{n}(2k)!}
        \right]^{1/2}
        \left(
        \begin{array}{ccc}
        k&n&k+n\\
        q&\mu&-q-\mu
        \end{array}
        \right)
        \frac{1}{R^{k+n+1}}\;M_{t\mu}^{(n)}\;C_{q+\mu}^{(k+n)}
        ({\bf R}_{t}/R_{t}),
        \end{eqnarray}
        \end{widetext}
where $\langle r^{k}\rangle$ is a mean value of $r^{k}$ for the
$5f$ orbital. ${\bf M}_{t}^{(n)}$ is the $2n$-pole electric
momentum induced on ion $t$, where $t$ runs over the all ions in
the infinite net and $\mu$ runs over the components of ${\bf
M}_{t}^{(n)}$. The state of electrostatic equilibrium between the
${\bf M}_{t}^{(n)}$ moment and the remaining multipole moments of
the crystal lattice is represented by the following equations
determining the electric multipole moments ${\bf
M}_{t}^{(n)}$:\cite{fg82,mg00}
        \begin{widetext}
        \begin{equation}\label{apbequi}
        {\bf M}_{t}^{(n)}=\sum_{t^{\prime}}\sum_{p=0,1,2}
        (-1)^{n+1}\;\alpha_{t}^{(n)} \;{\bf I}^{(2n)}\;
        {\bf \nabla}_{t^{\prime}}^{(n)}
        \left[
        {\bf M}_{t^{\prime}}^{(p)}\cdot
        {\bf \nabla}_{t^{\prime}}^{(p)}
        \frac{1}{{\bf R}_{t^{\prime}}}
        \right]
        \end{equation}
        \end{widetext}
where ${\bf I}^{(2n)}$ is the diagonal unit tensor of rank $2n$.
$\alpha_{t}^{(n)}$ is the $2n$-pole polarizability.

\section{\label{apc}}
Temperature dependence of the paramagnetic susceptibility is given
by the Van Vleck formula:\cite{v32}
\begin{eqnarray}\label{apcvv}
\chi_\alpha(T)=\frac{N_A\mu_B^2}{Z}\sum_{\gamma}(\beta a_{\gamma
,\alpha}+2b_{\gamma , \alpha})\exp(-\beta E_\gamma)
\end{eqnarray}
with
\begin{eqnarray}
a_{\gamma ,\alpha}=\sum_{\stackrel{\gamma^{\prime}}{
E_{\gamma^{\prime}}=E_{\gamma}}}\vert \langle \gamma \vert
L_{\alpha}+gS_{\alpha}\vert \gamma^{\prime} \rangle \vert
\end{eqnarray}
\begin{eqnarray}
b_{\gamma ,\alpha}=\sum_{\stackrel{ \gamma^{\prime}}{
E_{\gamma^{\prime}}\ne E_{\gamma}}} \frac{ \vert \langle \gamma
\vert L_{\alpha}+gS_{\alpha}\vert \gamma^{\prime} \rangle
\vert}{E_{{\gamma^{\prime}}} -E_{\gamma}}
\end{eqnarray}
\begin{eqnarray}
Z=\sum_{\gamma} \exp{(-\beta E_{\gamma)}}
\end{eqnarray}
$\beta =1/kT$, $\alpha= x, y, z$, index $\gamma$ numbers the
eigenstates, $E_{\gamma}$'s denote their energies, $L_{\alpha}$
and $S_{\alpha}$ are the $\alpha$ component of the total orbital
and spin operators and $g$ is the gyromagnetic ratio of the
electron spin.
\newpage
\bibliography{enh_gaj2}

\begin{thebibliography}{63}
\bibitem[{sta()}]{stand}
\bibinfo{note}{There is no a unique set of the $B_{kq}$ parameters in the case
  of low symmetry systems. Although, the parameders presented in the paper are
  strictly defined in the crystallographic coordination system, they have to be
  standardized before they are compared with other data. For further details
  see Refs.~\onlinecite{yeung,mg00,rq03,br04} and the references therein}.
\end{thebibliography}
\newpage
\end{document}